\documentclass[11pt]{article}

\usepackage[margin=1in]{geometry}

\usepackage[normalem]{ulem} 
\usepackage{siunitx}
\usepackage{upgreek}
\usepackage{graphicx}
\usepackage{amsmath}
\usepackage{amssymb}
\usepackage{xfrac}
\usepackage{soul}
\usepackage{pdfpages}
\usepackage{hyperref}
\usepackage{subcaption}
\usepackage{tabularx}
\usepackage{booktabs}
\usepackage{array}
\usepackage{bm}
\usepackage{mathptmx}

\usepackage[numbers,comma,sort&compress,square]{natbib}
\usepackage{titlesec}

\usepackage[font=footnotesize,labelfont=bf]{caption} 

\usepackage{float}
\usepackage[color=blue!15, size=footnotesize]{todonotes}
\usepackage{xcolor}

\titleformat{\section}{\Large\bfseries}{\thesection.}{1em}{}[]
\titleformat{\subsection}{\normalsize\bfseries}{\thesubsection.}{1em}{}[]
\titleformat{\subsubsection}{\normalsize\itshape}{\thesubsubsection.}{1em}{}[]
\titlespacing{\section}{0pt}{5pt}{0pt}[0pt]
\titlespacing{\subsection}{0pt}{3pt}{0pt}[0pt] 
\titlespacing{\subsubsection}{0pt}{0pt}{0pt}[0pt] 

\begin{document}

\begin{center}

\Large
\textbf{The structural relaxation time of a polymer glass during deformation}

\vspace{11pt}

\normalsize
Pradip K. Bera$^{1,\dagger}$, Grigori A. Medvedev$^{2}$, James M. Caruthers$^{2}$, Mark D. Ediger$^{1}$

$^1$ University of Wisconsin-Madison, Madison, Wisconsin 53706, United States\\
$^2$ Purdue University, West Lafayette, Indiana 47907, United States\\
\vspace{11pt}

$^\dagger$ Correspondence: pbera2@wisc.edu

\end{center}

\vspace{11pt}

\textbf{keywords:} polymer glass, structural relaxation, segmental correlation

\begin{abstract}
In order to determine the structural relaxation time of a polymer glass during deformation, a strain rate switching experiment is performed in the steady-state plastic flow regime. A lightly cross-linked poly (methyl methacrylate) (PMMA) glass was utilized, and simultaneously the segmental  motion in the glass was quantified using an optical probe reorientation method. After the strain rate switch, a non-monotonic stress response is observed, consistent with previous work. The correlation time for segmental motion, in contrast, monotonically evolves towards a new steady-state, providing an unambiguous measurement of the structural relaxation time during deformation, which is found to be approximately equal to the segmental correlation time. The Chen-Schweizer model qualitatively predicts the changes in the segmental correlation time and the observed non-monotonic stress response. In addition, our experiments are reasonably consistent with the material time assumption used in polymer deformation modeling; in this approach, the response of a polymer glass to a large deformation is described by combining a linear-response model with a time-dependent segmental correlation time. 
\end{abstract}

\section*{INTRODUCTION}
Polymer glasses play an important role in our modern life as flexible, transparent, light, and easily processable industrial materials. As glasses, their non-equilibrium nature is fundamental and the glass slowly evolves towards an equilibrium structure~\cite{hodge1995physical}. This slow structural evolution of the glass is limited by the rate of molecular rearrangements, and occurs at a rate that depends on the temperature, pressure, chemical composition, and thermobaric history~\cite{ediger1996supercooled}. ``Physical aging'' describes the changes in physical properties that occur as result of this evolution, for example, the viscoelastic shear modulus typically increases during aging~\cite{struik1978physical}. Notably, physical aging does not involve any chemical change, and it can be completely reversed by heating the sample above the glass transition temperature, in a process known as thermal rejuvenation. Even with no temperature change, other external stimuli such as shear or tensile deformation can also reverse (some of) the effects of physical aging~\cite{parmar2019strain,mota2021enhancing}. At a qualitative level, one can envision the physical aging of glass as the system moving lower on the potential energy landscape (PEL) where the barriers to rearrangement are higher, while thermal or mechanical rejuvenation can be envisioned as moving up the PEL to where the barriers are lower~\cite{mckenna2003mechanical,yang2022role,riggleman2008nonlinear,liu2010aging}.

One of the most important quantities for understanding a glass-forming system is the structural relaxation time~\cite{kovacs1964transition,scherer1986relaxation,malek2001structural,priestley2005structural,parmar2017density,adhikari2023dependence}. To obtain this quantity, often a small temperature jump (either positive or negative) is applied to drive the system from the equilibrium liquid at T$_1$ to the equilibrium liquid at T$_2$~\cite{richert2014supercooled,richert2022one}. It has been established that the density, modulus, and molecular mobility of a glass all change monotonically in response to such a temperature jump, and all these quantities reach their new equilibrium values at about the same time~\cite{hecksher2015communication,hecksher2019fast}. To a reasonable approximation, the equilibration time for any of these quantities can be used to determine the structural relaxation time, which can be envisioned as the characteristic time to move between two levels on the PEL~\cite{ekimoto2013theoretical}. Recent work has made extensive comparisons between the structural relaxation time (measured through a small temperature jump) and underlying molecular motion in the equilibrium liquid,~\cite{hecksher2010physical,schawe2014vitrification,saini2018interplay,richert2022one,di2023physical,lapuk2023some} with the conclusion that in absence of other external stimuli such as shear or tensile deformation, the primary ($\alpha$) process of the equilibrium liquid controls the observed structural relaxation, at least so long as the temperature jump is small. In qualitative terms, for quiescent polymeric systems, this establishes that segmental rearrangements control the physical aging process.

Here, we measure the structural relaxation time of a polymer glass $\textit{during}$ $\textit{tensile}$ $\textit{deformation}$. When a polymer glass is subjected to mechanical forces, as happens in many critical applications, the structure of the glass and the position on the PEL can evolve in response. Many theoretical efforts attempt to describe this complex process, often in the form of constitutive equations. As all current approaches appear to have deficiencies,~\cite{anand2006modeling, van2011extending} we attempt here to advance our understanding through an analogy with the temperature jump experiments described above. Generalizing, we define the structural relaxation time of a polymer glass during deformation as the time required for the isothermal transition from one mechanical steady-state to another. Nanzai and coworkers performed such an experiment on a PMMA glass,~\cite{nanzai1990transition} making use of the plastic flow regime that occurs after yield during mechanical deformation at a constant strain rate. Once a steady-state (constant stress) flow had been established, they quickly switched the strain rate. Interestingly, a stress undershoot appeared when the strain rate was decreased by two orders of magnitude. Unfortunately, this non-monotonic response of the stress creates significant ambiguity for a determination of the structural relaxation time. Furthermore, there is no consensus on the interpretation of this striking stress undershoot. Recently, Medvedev and Caruthers investigated a series of theoretical models and found that several models failed to produce the stress undershoot after strain rate switching, while other models could qualitatively reproduce the experimental observations~\cite{medvedev2016comparison}.

Here we revisit the Nanzai's strain rate switching experiment with a new tool: the measurement of the segmental correlation time $\tau_{seg}$ during deformation. Recent experiments and simulations have shown that the segmental dynamics of the polymer glass are intimately connected with the deformation response~\cite{lee2008dye,lee2009deformation,perez2016dielectric,sahli2020relaxation}. As the polymer glass yields and flows, the segmental correlation time is reduced by an order of magnitude or more. Qualitatively, deformation turns a solid into a liquid, by increasing the rate of segmental relaxation until it is comparable to the deformation rate. This fundamental connection between the polymer segmental correlation time and deformation is embedded in almost all theoretical models through the use of the concept of material time~\cite{narayanaswamy1971model,riechers2022predicting}. With this approach, the response of a polymer glass to a large deformation can be captured by the model describing the linear response regime, except that the segmental correlation time becomes a time-dependent variable that is altered by the deformation~\cite{riechers2022predicting}. Though widely used, the validity of the material time approximation is unclear.

\section*{RESULTS}

\begin{figure}
\includegraphics[width=1.0\textwidth]{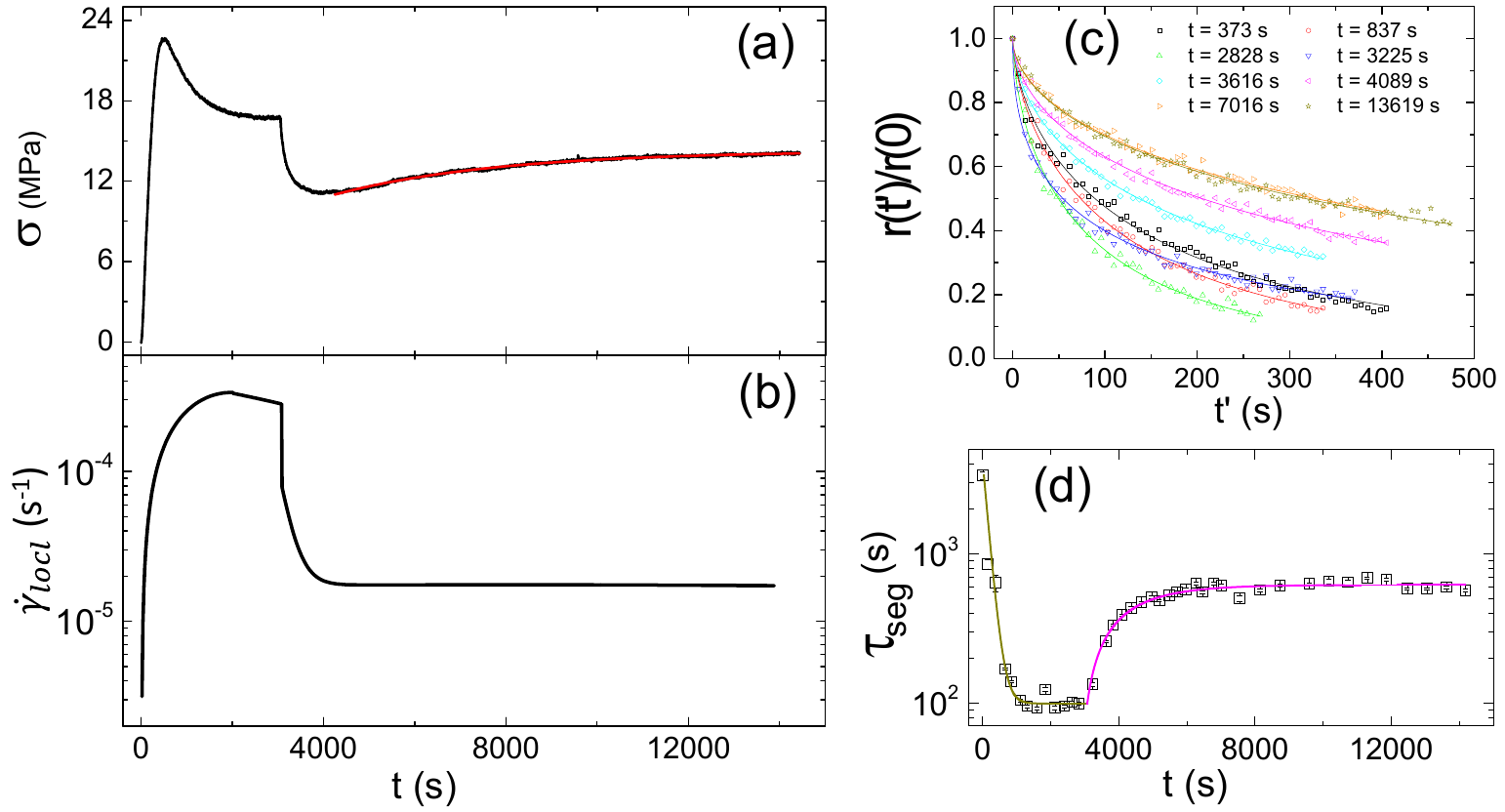}
\caption{High to low strain rate switching experiment (switching from $6 \times 10^{-5}~s^{-1}$ to $6 \times 10^{-6}~s^{-1}$ at $t = 3075~s$). (a) Time dependence of measured stress $\sigma$; along with the functional fit $Y = Y_s + (Y_i - Y_s)exp(-(t-t_0)/\tau_Y)$ starting after the minimum at $4000~s$, shown as the red curve. The fitted transition time is $\tau_{\sigma} = 3760 \pm 160~s$. (b) Measured local strain rate; $\dot{\gamma}_{locl}$ is plotted with $t$. (c) A subset of the normalized anisotropy decay curves $r(t')/r(0)$, measured at various times $t$ after the start of deformation, are shown as examples. Solid lines are KWW fits. (d) The segmental correlation time $\tau_{seg}$ extracted from the KWW fits are plotted vs time $t$. Error bars show fitting errors. Solid curves are similar functional fits as in (a) with $\tau_{\tau-seg} = 165 \pm 10$ for the green curve, and  $\tau_{\tau-seg} = 1330 \pm 110$ for the magenta curve.}
\label{F1}
\end{figure}

In this study, we mechanically induce an isothermal, steady-state to steady-state transition of the PMMA polymer glass at $380~K$ i.e. T${_g} - 19~K$. We apply a constant strain rate to deform the PMMA glass and wait until it reaches steady-state in the flow regime. Then, we impose an order of magnitude switch to the applied strain rate and wait again for the material response to reach the new steady-state. We follow the local strain rate ($\dot{\gamma}_{locl}$) and the segmental dynamics in situ, in order to measure the structural relaxation time of the polymer glass as it responds to the strain rate switch. Additional experimental details may be found in the Supplemental Material~\cite{mysupmat2024prl}. The anisotropy decay after photobleaching due to the reorientation of the DPPC probe has previously been shown to be an excellent reporter of the segmental dynamics of PMMA~\cite{lee2009direct,bending2014measurement,ricci2018direct}. The time scale $\tau_{seg}$ extracted by Kohlrausch-Williams-Watts (KWW) function fitting to the anisotropy decay is a measure of the time scale of the PMMA segmental dynamics. 

We first describe the high-to-low strain rate switching experiment. We set $t = 0~s$ at the start of sample deformation, i.e., after the $36~min$ aging. We initially applied the global strain rate $\dot{\epsilon} = 6 \times 10^{-5}~s^{-1}$ until $t = 3075~s$, and then we switched to $\dot{\epsilon} = 6 \times 10^{-6}~s^{-1}$. The measured global stress $\sigma$ ($= \frac{measured~force}{2.3~ mm \times 50~ \mu m}$) is plotted against the duration of the experiment in Fig.~\ref{F1}a. We can observe an initial linear region due to elasticity, then yielding with maximum $\sigma \approx 22.6~MPa$, followed by strain softening, and then a steady-state ($\approx 17~MPa$) until $3075~s$. After switching the strain rate, $\sigma$ shows a minimum around $4000~s$, and then goes to the second steady-state $\approx 14.5~MPa$. The extracted time scale for $\sigma$ to reach the steady-state after the minimum is $\tau_{\sigma} = 3760 \pm 160~s$. The observed behavior of $\sigma$ following the strain rate drop is in qualitative agreement with Nanzai's compressive experiment with PMMA~\cite{nanzai1990transition}. The measured local strain rate $\dot{\gamma}_{locl}$ does not immediately switch to the lower value. Instead, it takes $\sim 1000~s$ (Fig.~\ref{F1}b) to make a monotonic transition from the first steady-state strain rate ($\approx 3 \times 10^{-4}~s^{-1}$) to the second steady-state strain rate ($\approx 2 \times 10^{-5}~s^{-1}$). We stopped this experiment at $14000~s$. We then reversed the linear actuator back to the initial position and thermally rejuvenated the sample to erase history, enabling further measurements with the same sample. 

We repeated this high-to-low strain rate switching experiment two additional times, in order to perform fluorescence anisotropy measurements at several time points during the transition between steady-states. The global stresses for these three experiments are in excellent agreement, confirming the repeatability of the experiment (Supplemental Material Fig. S2)~\cite{mysupmat2024prl}. A subset of the anisotropy decay curves are shown in Fig.~\ref{F1}c, along with fits to the KWW function. The extracted segmental correlation time ($\tau_{seg}$) from each anisotropy curve is plotted against deformation time ($t$) in Fig.~\ref{F1}d, at the time corresponding to the mid-point of the anisotropy decay measurement. We observe an order of magnitude decrease in $\tau_{seg}$ (from $\approx 3300~s$ to $\approx 100~s$) due to yielding. After strain rate switching, there is a monotonic increase in $\tau_{seg}$ to reach the second steady-state ($\tau_{seg} \approx 620~s$). In Fig.~\ref{F1}d, we show fitted curves that reproduce the variations of $\tau_{seg}$ with two different functions used before and after switching. For the steady-state to steady-state transition, the extracted time scale for the change in $\tau_{seg}$ is $\tau_{\tau-seg} = 1330 \pm 110~s$ which is significantly $< \tau_{\sigma}$, the time scale of $\sigma$. 
 
\begin{figure*}
\includegraphics[width=1.0\textwidth]{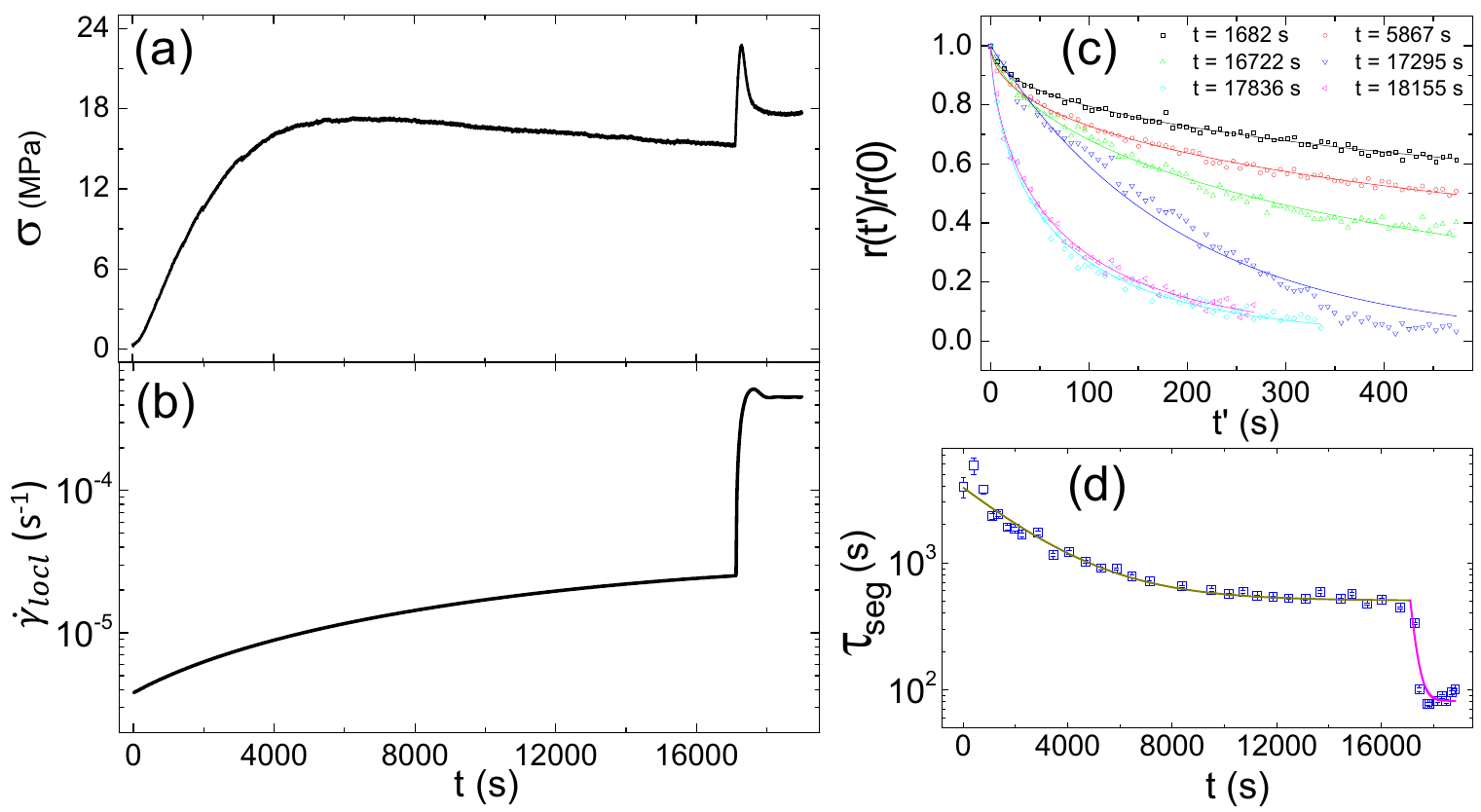}
\caption{Low to high strain rate switching experiment (switching from $6 \times 10^{-6}~s^{-1}$ to $6 \times 10^{-5}~s^{-1}$ at $t = 17114~s$). Variations of $\sigma$, $\dot{\gamma}_{locl}$, a few $r(t')/r(0)$, and $\tau_{seg}$ are shown in (a), (b), (c), (d) respectively. Error bars are fitting errors. Solid curves in (d) are the functional fits $Y = Y_s + (Y_i - Y_s)exp(-(t-t_0)/\tau_Y)$ with time scales $\tau_{\tau-seg} = 2365 \pm 85~s$ for the green curve and $\tau_{\tau-seg} = 200 \pm 45$ for the magenta curve.}
\label{F2}
\end{figure*}

With this same sample, we have also performed low to high strain rate switching experiment, switching from $\dot{\epsilon} = 6 \times 10^{-6}~s^{-1}$ to $\dot{\epsilon} = 6 \times 10^{-5}~s^{-1}$. In this case, $\sigma$ shows an overshoot at $t = 17300~s$ (Fig.~\ref{F2}a) having very nearly the same maximum value as in Fig.~\ref{F1}a before transitioning to the second steady-state at $t \sim 17800~s$. The local strain rate ($\dot{\gamma}_{locl}$) goes to the second steady-state at $t \sim 17900~s$ (Fig.~\ref{F2}b). Again a subset of the anisotropy decay curves measured during this experiment are shown in Fig.~\ref{F2}c. The decay curve observed immediately after switching cannot be well described by the KWW function; this will be discussed below. When the strain rate suddenly switched, the extracted $\tau_{seg}$ values decrease and then become steady at $t \sim 17800~s$ (Fig.~\ref{F2}d). In this case, all the three measured quantities $\sigma$, $\dot{\gamma}_{locl}$, and $\tau_{seg}$ reach the steady-state so quickly that we cannot meaningfully distinguish their transition times. Fits capturing the variations in $\tau_{seg}$ are shown before ($\tau_{\tau-seg} = 2365 \pm 85~s$) and after ($\tau_{\tau-seg} = 200 \pm 45~s$) the switching.

During the strain rate switching experiments, there are significant changes in the KWW non-exponentiality parameter $\beta$~\cite{ediger2000spatially}. For a system at steady-state, $\beta$ can be interpreted in terms of a distribution of segmental correlation times (the characteristic length scale for dynamic heterogeneity is a few nanometers). $\beta$ has limits $0< \beta <1$, where smaller values of $\beta$ indicate a broader distribution of correlation times. Deformation has been previously reported to narrow the distribution of the segmental correlation time within a polymer glass~\cite{lee2009deformation,lee2010mechanical}. Here, for both types of strain rate switching experiments, $\beta$ increases from 0.4 to a steady value in range of 0.6 to 0.7, depending upon the applied strain rate (Supplemental Material Fig. S3)~\cite{mysupmat2024prl}. This is consistent with previous studies~\cite{hebert2015effect}. After the strain rate switching, $\beta$ transitions from one steady-state value to another, with anomalous KWW $\beta$ parameters being obtained immediately after the switch. These anisotropy curves are not well-described by the KWW function and are further discussed below.

\begin{figure}
\includegraphics[width=1.0\textwidth]{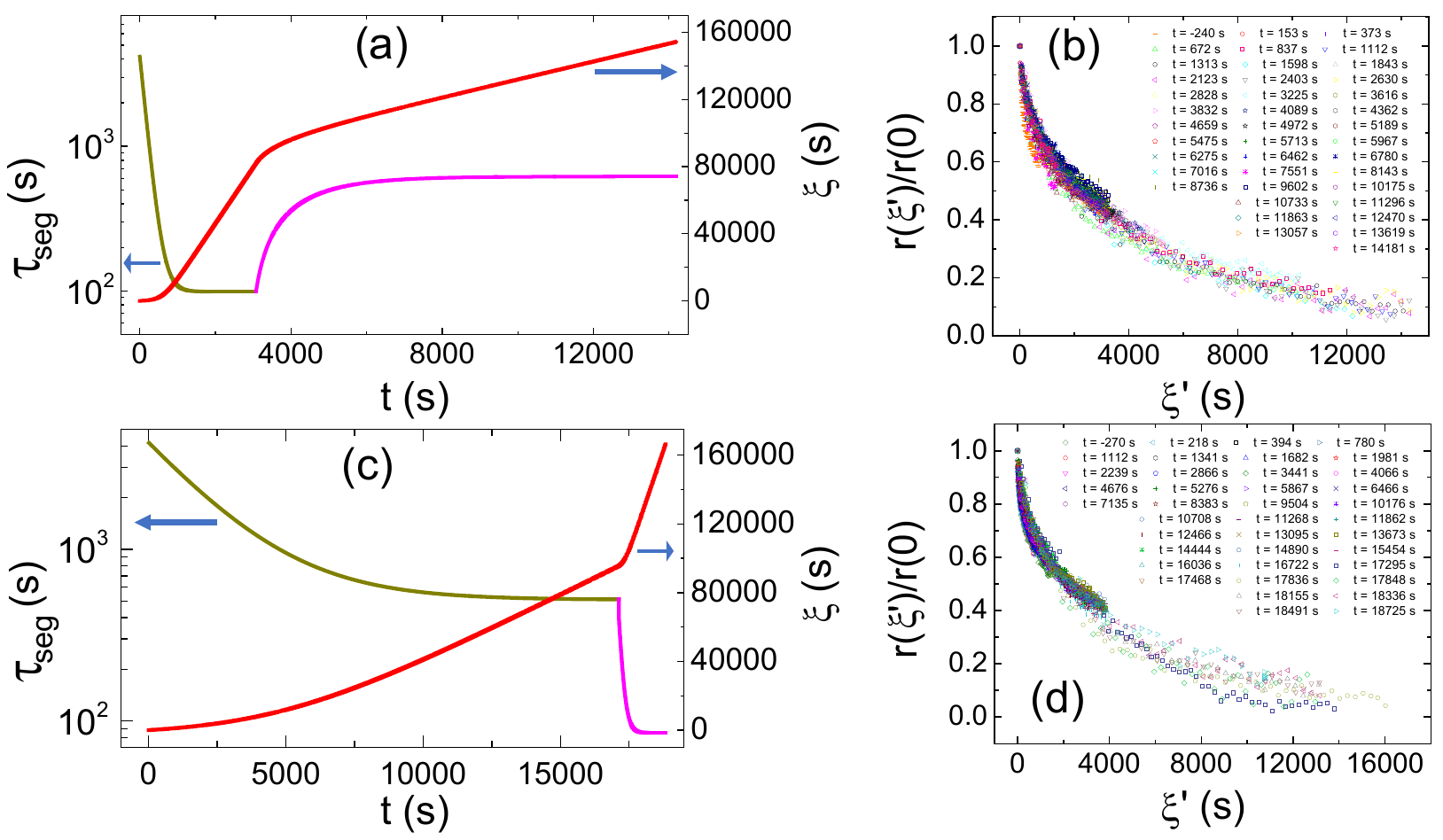}
\caption{Experimental time vs. material time: (a and c) the variation of $\tau_{seg}$ vs $t$ during the full span of the experiments (fitted data from Fig.~\ref{F1}(d) and Fig.~\ref{F2}(d)) is shown along with the calculated material time. (b and d) Normalized anisotropy decay curves for the two types of experiments are re-plotted. In material time, $\xi'$ is used to denote elapsed material time since the beginning of an anisotropy measurement.}
\label{F3}
\end{figure}

In order to further analyze our results, we test whether our anisotropy decay curves are consistent with the material time approximation (i.e. $\beta \sim$ constant)\cite{narayanaswamy1971model,di2023physical}. Models that make this approximation typically couple the mechanical response expected in the linear response regime with a variable clock rate that accounts for the effects of non-linear deformation. If this approximation is valid, the anisotropy decay functions which show substantial variation in Figures~\ref{F1}c and~\ref{F2}c, would all superpose, when viewed as a function of material time. We use the observed time variations of $\tau_{seg}$ i.e. the fitted curves in Fig.~\ref{F1}d and Fig.~\ref{F2}d, to calculate the variation of material time ($\xi$) with respect to the real-time ($t$) using $\xi = \int_{0}^{t}\frac{\tau_u}{\tau_{seg}}~dt'$, where $\tau_u$ is the value of $\tau_{seg}$ before starting the deformation ~\cite{di2023physical}. In Fig.~\ref{F3}(a), and Fig.~\ref{F3}(c), $\xi$ is plotted with respect to $t$ respectively.

Remarkably, when we look at all the anisotropy decay curves in terms of material time, all the decay curves overlap well, including the decay curve just before deformation (Fig.~\ref{F3}b, and Fig.~\ref{F3}d). This agreement also includes the decay curves observed immediately after strain rate switching which were poorly fit using KWW function (Supplemental Material Fig. S4)~\cite{mysupmat2024prl}. The non-linear variation of material time shifts the data points of decay curves in such a way that all the decay curves fall close to the undeformed one. As discussed above, material time is a common assumption in models of polymer deformation and these results indicate that it is a reasonable approximation for the deformations considered here.

\begin{figure}
\includegraphics[width=1.0\textwidth]{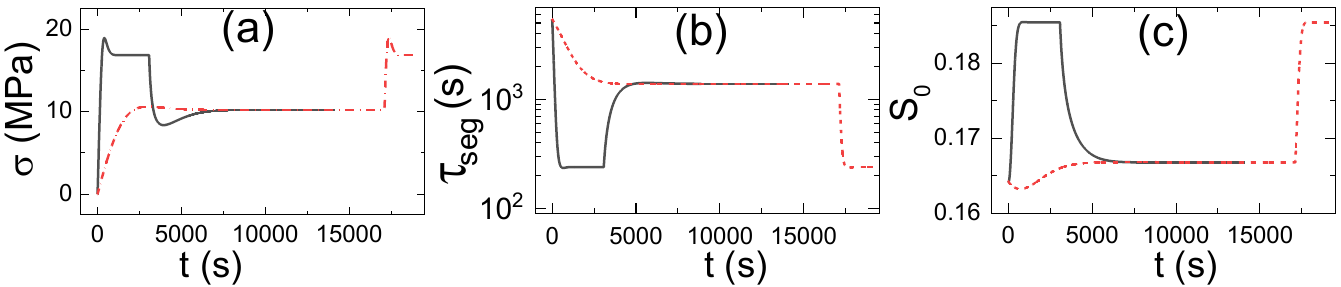}
\caption{Chen-Schweizer model predictions for strain rate switching experiments on PMMA glass, using the global strain rates $6 \times 10^{-5}~s^{-1}$ and $6 \times 10^{-6}~s^{-1}$ as inputs. Model outputs are the time variation of (a) the stress $\sigma$, (b) the segmental correlation time $\tau_{seg}$, and (c) the structural state variable $S_0$. Black curves for high to low strain rate switching, and dotted red for low to high strain rate switching.}
\label{F4}
\end{figure}

We can now address the following questions regarding our high to low strain rate switching data: (1) Can we determine and understand the structural relaxation time for the PMMA glass for the transition between the two steady-state flow regimes? (2) Why does $\sigma$ show an undershoot after switching to a low strain rate, whereas the segmental correlation times $\tau_{seg}$ increase monotonically? To address these questions, we compare our data to the nonlinear Langevin equation theory proposed by Chen and Schweizer~\cite{chen2007molecular, chen2008microscopic,chen2010theory,chen2011theory}. In this approach, a structural state variable $S_0$, related to the amplitude of density fluctuations on the nanometer scale, controls the response of the glass to deformation.   Fig.~\ref{F4}a-c show the predictions of the Chen-Schweizer model for the strain rate switching experiment for $S_0$, the stress $\sigma$, and the segmental correlation time $\tau_{seg}$ (see Supplemental Material for details about the model calculations)~\cite{mysupmat2024prl}. All of the qualitative features of the experiments are well-described by the model calculations. In particular, the predicted variations of both $\sigma$ and $\tau_{seg}$ resemble the experimental observations of strain rate switching. We have extracted the characteristic relaxation times $\tau_{\sigma}$, $\tau_{\tau -seg}$, $\tau_S$ associated with the variations of the parameters $\sigma$, $\tau_{seg}$, and $S_0$ respectively (Supplemental Material Fig. S5)~\cite{mysupmat2024prl}, in a manner analogous to our treatment of the experimental data. 

To address the first question, regarding experimental access to the structural relaxation time for the transition from one steady-state flow to another, we observe that $\tau_{\tau -seg}  \approx \tau_S$. This was also noted by Chen and Schweizer~\cite{chen2010theory} and within the context of the model, $\tau_S$ is the structural relaxation time. So the experimental observation of $\tau_{\tau-seg}$ provides a reasonable estimate of the $\tau_S$ in case of steady-state to steady-state transition of flow of polymer glass. To verify the generality of this conclusion, we performed calculations with a less complex model (``Toy model 1'' from ref.~\cite{medvedev2022multistep}). Similar to the Chen-Schweizer model, we observed that the state variable and $\tau_{seg}$ approached steady-state on a similar time scale. Furthermore, both our results and the Chen-Schweizer model indicate that the structural relaxation time during deformation is very similar to the segmental correlation time itself. Thus our work extends the connection between segmental mobility and structural relaxation from the quiescent state (where it is well established) to a mechanically driven steady-state.

With regard to the second question, we can address why $\sigma$ shows an undershoot after switching to a low strain rate, whereas the segmental correlation time $\tau_{seg}$ increases monotonically. The model successfully reproduces the experimental observation of slower variation of $\sigma$ compared to $\tau_{seg}$ i.e. $\tau_{\sigma} > \tau_{\tau-seg}$. To get a detailed understanding of the undershoot in $\sigma$, we have done additional numerical calculations with the model equations in which we keep the $\tau_{seg}$ and $S_0$ precisely fixed after the $\tau_{seg}$ has very nearly reached its steady-state value. Under these conditions, $\sigma$ continues to vary for a considerable length of time, mainly controlled by the second term of the generalized Maxwell constitutive equation in Chen and Schweizer model~\cite{chen2011theory} i.e. $-\frac{\sigma}{\tau_{seg}}$ as $\sigma << \sigma_c$ (Supplemental Material Equation~\ref{eqn1}, Fig. S6)~\cite{mysupmat2024prl}. This lag in the stress response was also pointed out by Nanzai~\cite{nanzai1990transition}. Within the Chen and Schweizer model, this interpretation is possible: the undershoot occurs when a system high in the landscape (as a result of fast deformation) suddenly experiences slower deformation; the stress naturally drops when the strain rate goes down, but it takes some time (roughly $\tau_S$) for the system to “age” down to its new position in the landscape. As the system ages, the barriers to relaxation grow, and thus the stress must increase in order to maintain the strain rate.

We see two important directions for extending this work. Similar experiments further below T${_g}$ would test the generality of our results. Recent work~\cite{hebert2015effect} has shown that the segmental correlation time of a deformed polymer glass in the flow regime is a weak function of temperature, for a given strain rate. We anticipate that, at lower temperatures, the structural relaxation time during deformation continues to be controlled by the (deformation-accelerated) segmental mobility. Together, these two observations suggest that strain rate switching experiments further below T${_g}$ will be qualitatively similar to the ones shown here. Additionally, there is a need for better models to describe the deformation of polymer glasses. Most models of polymer glass deformation do not describe the distribution of correlation times known to be important in glassy systems~\cite{medvedev2016comparison}. The Chen-Schweizer approach has recently been extended to account for heterogeneous dynamics~\cite{ghosh2020role}. The increase of the KWW $\beta$ parameter with increasing strain rate (as seen here and previously~\cite{hebert2015effect}) is at least qualitatively accounted for in the new work.

\section*{SUMMARY}

To measure the structural relaxation time for a polymer glass during deformation, we have performed strain rate switching measurements on a PMMA glass at T$_g - 20~K$. In addition to observing the resulting stress, we have measured the time variation of the segmental correlation time of the polymer. We have shown that the correlation functions describing segmental mobility are approximately invariant when plotted in material time. This result supports the use of material time in models of polymer glass deformation. The Chen-Schweizer model qualitatively describes the evolution of the stress and segmental correlation time, for the strain rate switching experiments. It allows an interpretation of the undershoot of the stress in terms of the energy landscape. Our optical experiments indicate that the structural relaxation time during deformation is very similar to the segmental correlation time itself. Thus our work extends the connection between segmental mobility and structural relaxation from the quiescent state to a mechanically driven steady-state.

\section*{ACKNOWLEDGMENTS}
G.A.M. and J.M.C. acknowledge support from National Science Foundation Grant No. 1761610-CMMI. P.K.B. and M.D.E. acknowledge support from the National Science Foundation (DMR) grant number DMR-2002959.


\clearpage


\setcounter{page}{1}
\pagenumbering{arabic}

\setcounter{figure}{0}
\setcounter{table}{0}
\setcounter{equation}{0}
\setcounter{section}{0}

\renewcommand{\thefigure}{S\arabic{figure}}
\renewcommand{\thetable}{S\arabic{table}}
\renewcommand{\theequation}{S\arabic{equation}}

\renewcommand{\figurename}{Supplementary Fig.}


\begin{center}

\Large
\textbf{Supplemental Material: The structural relaxation time of a polymer glass during deformation}

\vspace{11pt}

\normalsize
Pradip K. Bera$^{1,\dagger}$, Grigori A. Medvedev$^{2}$, James M. Caruthers$^{2}$, Mark D. Ediger$^{1}$

$^1$ University of Wisconsin-Madison, Madison, Wisconsin 53706, United States\\
$^2$ Purdue University, West Lafayette, Indiana 47907, United States\\
\vspace{11pt}

$^\dagger$ Correspondence: pbera2@wisc.edu

\end{center}

\vspace{11pt}

\textbf{keywords:} polymer glass, structural relaxation, segmental correlation

\section*{EXPERIMENTAL DETAILS}
\begin{figure*}
	\includegraphics[width=1.0\textwidth]{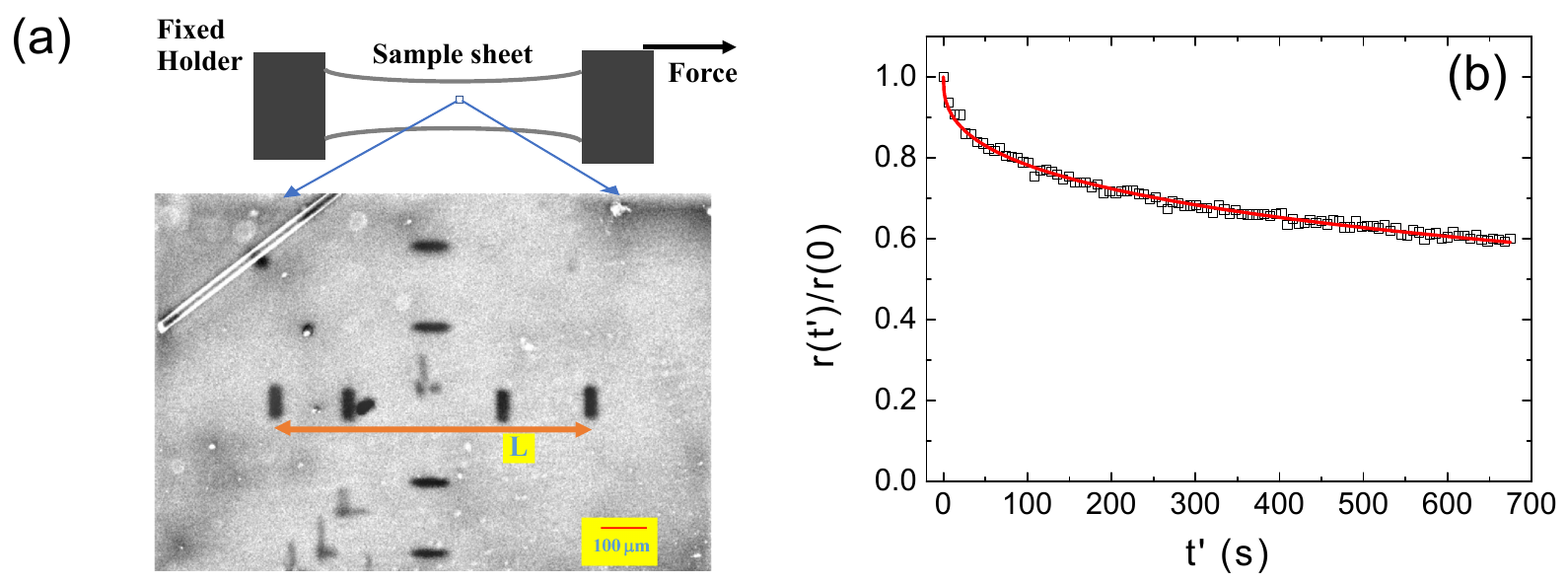}
	\caption{Study of PMMA glass during deformation: (a) schematic of the polymer glass sheet being deformed, and fluorescence image of the glass sheet at $380~K$ (doped with $5~\mu M$ DPPC as optical probe). The vertical straight lines are used for local strain rate measurements. The `L'-like patterns were used to measure the fluorescence anisotropy, produced by the polarized bleaching beam. (b) For the undeformed sample, graph of the normalized anisotropy $r(t')/r(0)$ vs $t'$, providing access to the segmental correlation time of the polymer. Red line is the KWW-fit.}
	\label{SIF1}
\end{figure*}

In this study, we have used sheets of lightly cross-linked poly(methyl methacrylate) (PMMA) as a model glass. Light cross-linking is used to increase the ductility of the sample and to allow repeated measurements on the same sample. Light cross-linking has a slight effect on T${_g}$ (a few K) but properties of the glass are essentially unchanged in comparison to the uncross-linked material at the same T – T${_g}$. Physical aging properties in the linear response regime, for example, are indistinguishable when compared in this way~\cite{siricci2018direct}. Lightly cross-linked PMMA polymer glass was synthesized in our lab. For our strain rate switching experiments, we have used a $21~mm \times 2.3~mm \times 50~\mu m$ PMMA glass sheet (effective length after end-clampings $\approx 16.4~mm$) doped with $\approx 5 \times 10^{-6}~M$ N,N$^\prime$-dipentyl-3,4,9,10-perylenedicarboximide (DPPC) as an optical probe (Fig.~\ref{SIF1}a). To prepare the PMMA glass we have used $98.5~wt\%$ MMA, with $1.5~wt\%$ ethylene glycol dimethacrylate as a crosslinker, and $\approx 5 \times 10^{-6}~M$ of DPPC as optical probes. The initiator $0.1~wt\%$ benzoyl peroxide was added to the stock solution to start polymerization. The mixture was placed in a water bath at $345~K$ to do initial polymerization for $30~min$ and then transferred to molds, which were made with two $2 \times 3~in$ glass slides separated with aluminum foil (for desired thickness) and clamped with binder clips. The molds were kept in a nitrogen environment for $24~hours$ at $345~K$ for further polymerization. After that PMMA films were removed from molds by sonication and were cut into dog-bone shaped pieces with a custom die. After loading, the samples were annealed at $420~K$ under nitrogen for $24~hours$. A detailed description of the sample preparation method can be found in previous publications~\cite{silee2009direct}. The glass transition temperature of the sample, T$_g = 399 \pm 1~K$, was determined from the midpoint of the glass transition region in the second DSC scan at $10~K~min^{-1}$. Before each experiment, the sample was annealed at $420~K$ for $30~min$ to erase thermal and deformation history, and then cooled at a rate $2~K~min^{-1}$ to the measurement temperature $380~K$. After 36 minutes of aging, the deformation experiment was started. All the deformation experiments reported here were performed at $380~K$ i.e. T${_g} - 19~K$.

Using a programmable linear actuator, we applied an extensional stress by pulling one end of the PMMA glass sample inside a temperature-controlled chamber, resting on a fluorescent microscope, while measuring the force with a load cell (Fig.~\ref{SIF1}a). To apply a strain rate deformation, we apply actuator speed $0.001~mm~s^{-1}$ for the global strain rate $\dot{\epsilon} = 6 \times 10^{-5}~s^{-1} (= 0.001~mm~s^{-1}/16.4~mm)$, or an applied speed $0.0001~mm~s^{-1}$ for $\dot{\epsilon} = 6 \times 10^{-6}~s^{-1}$. For local strain rate measurements, we have grabbed ($0.1~fps$) fluorescent images of the straight lines perpendicular to the direction of deformation. The images were further analyzed using ImageJ software~\cite{sischindelin2012fiji} to get the local strain $\gamma_{locl} = \frac{L(t)-L_0}{L_0}$ where $L_0$ and $L(t)$ are the horizontal distance between the drawn lines before deformation and at time t during deformation, respectively. Instantaneous local strain rates ($\dot{\gamma}_{locl}$) during deformation are determined by taking the derivative of a polynomial fit to the local strain versus time data as described in a previous publication~\cite{sibending2014measurement}. Within the field of view of our microscope ($\sim 700~\mu m~\times~700~\mu m$), we observe homogeneous deformation on the scale of microns as reported earlier~\cite{silee2009direct}. The maximum global strain that can be applied to samples of this type is $\sim 0.3$, for the conditions of our experiments. Here, we have limited the global strain $\sim 0.2$ in order to use one sample for all our measurements.

The segmental dynamics of the PMMA glass before and during deformation were measured in-situ using fluorescence recovery after the photo-bleaching (FRAP) technique (see ref.\cite{silee2009direct} for the technique details). A small concentration ($\approx 10^{-6}~M$) of randomly oriented fluorescent probe molecules (DPPC) is present in the PMMA glass sample. During each FRAP measurement, a small area of the sample is exposed to an intense linearly polarized laser beam ($532~nm$) for about $20~s$ to photo-bleach some DPPC molecules having transition dipoles parallel to the polarization direction. A circularly polarized reading beam ($532~nm$) is employed to induce fluorescence from the bleached region, which is collected and passed through a polarizing beam-splitter, allowing for the independent measurements of fluorescence parallel and perpendicular to the polarization of the bleaching beam. The orientational anisotropy of fluorescence, $r(t')$ (the time counter $t'$ starts from 0 as soon as the first data point $r(0)$ appears) is calculated from the difference between the parallel and perpendicular components of fluorescence. As the measurement proceeds, motion of the polymer segments leads to probe reorientation, causing the $r(t')$ to decrease. The $r(t')/r(0)$ is fitted with the Kohlrausch-Williams-Watts (KWW) function, $r(t') = r(0)exp (-(t/\tau_{seg})^\beta)$ to extract the segmental correlation time $\tau_{seg}$. In Fig.~\ref{SIF1}b, we show a representative anisotropy decay curve $r(t')/r(0)$ vs $t'$ which was measured just before deforming the sample; here the quantity $t'$ denotes the time elapsed since the beginning of an anisotropy decay measurement.

\clearpage
\section*{Chen and Schweizer model calculations}
 The molecular Eyring theory describes segmental mobility, elasticity, and nonlinear mechanical response of polymer glass, including the interplay between aging and deformation. In this model, the stress $\sigma$, the structural state variable $S_0$, and the segmental correlation time $\tau_{seg}$ depend on the strain rate ($\dot{\gamma}$) in a coupled way as given by equations (~\ref{eqn1}-\ref{eqn3}).
\begin{equation} \label{eqn1}
	\frac{d\sigma}{dt} = 2.8 G_0 (1-\frac{2}{3}\frac{\sigma}{\sigma_c})\dot{\gamma} - \frac{\sigma}{\tau_{seg}} \,,\,\,\,\,\,\,
\end{equation}
\begin{equation} \label{eqn2}
	\frac{dS_0}{dt} = \psi (S_g-S_0)\sigma\dot{\gamma} - \frac{(S_0-S_l)}{\tau_{seg}(S_0, \sigma = 0)} \,,\,\,\,\,\,\,
\end{equation}
\begin{equation} \label{eqn3}
	\tau_{seg} = \tau_0exp(0.4(\lambda - \lambda_c)^{1.3}(1-\frac{|\sigma|}{\sigma_c})^{2.5})  \,,\,\,\,\,\,\,
\end{equation}
where $\lambda = S_0^{-1.5}$, and $G_0$, $\sigma_c$, $\psi$, $S_g$, $S_l$, $\tau_0$, $\lambda_c$ are model parameters. Here, we have taken $\psi = 1$ and $\lambda_c = 8.32$ as in ref.~\cite{sichen2010theory}. We use the initial slope in main text Fig.1a to get $G_0 \approx 470~MPa$. To fix other parameters in these equations, we have compared the theory to aging data for a PMMA glass~\cite{siricci2018direct}. For a pure aging system, equation~\ref{eqn1} will vanish, the first part of equation~\ref{eqn2} will be absent, and the second factor in equation~\ref{eqn3} equals to unity. We can reproduce the observed aging data~\cite{siricci2018direct} by setting $S_l = 0.148$, $\tau_0 = 47$, and $S_0 = 0.243$ at $t=0$. Since the PMMA samples used here were aged prior to deformation, we use the value of $S_0$ at $t=2160~s$, as the initial value for our samples, just prior to deformation. Finally, we optimized the values $\sigma_c = 100~MPa$ and $S_g = 0.245$ to get the best prediction for the strain rate switching experiments. Note that we have used the global strain rates as inputs to the model.

\clearpage
\section*{Repeatability of experiments with PMMA}
\begin{figure}[!h]  
\begin{center}
\includegraphics[width=1.0\textwidth]{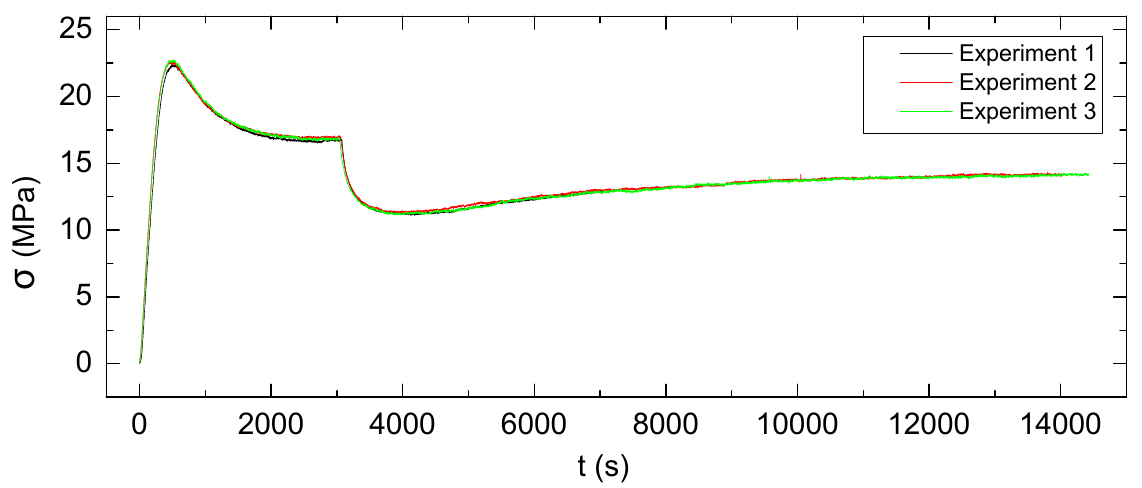}\\
\caption{Combined plot of $\sigma$ vs $t$ for three high to low strain rate switching experiments, switching at $t = 3075~s$ from the global strain rate $6 \times 10^{-5}~s^{-1}$ to $6 \times 10^{-6}~s^{-1}$.}
\label{SIF2}
\end{center}
\end{figure}

 We have checked the repeatability of our experiments with the PMMA sample. With our annealing protocol as described below, we can completely erase the thermal and deformation history of the PMMA sample. Before each experiment, the sample was annealed at $420~K$ for $30~min$ and then cooled to the measurement temperature $380~K$ at a rate $2~K~min^{-1}$. After that, we wait for 36 minutes to age the sample, and then we start the deformation experiment. To compare, we have plotted three sets of stress data ($\sigma$) for the three high-to-low strain rate switching experiments (Fig.~\ref{SIF2}) - the fluorescence microscopy experiment (Experiment 1), the fluorescence anisotropy experiment (Experiment 2), another fluorescence anisotropy experiment to do decay measurement in intermediate times (Experiment 3). A good overlap of the three sets of data indicates that we can repeat our experiments many times as our annealing protocol is an excellent one to remove the sample history.

\section*{Variation of $\beta$ during strain rate switching experiment}
\begin{figure}[!h]
\begin{center}
\includegraphics[width=0.9\textwidth]{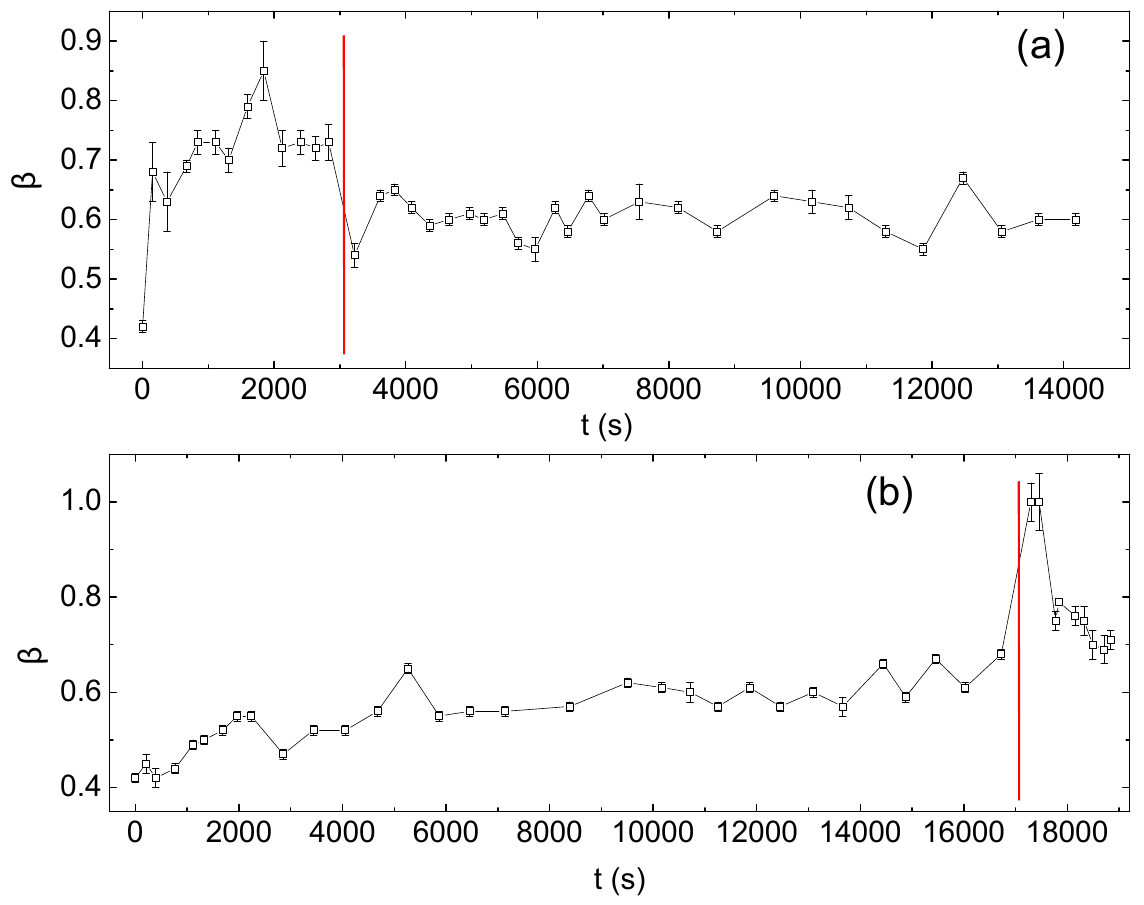}\\
\caption{KWW fitting parameter $\beta$ is plotted vs $t$ for (a) high to low and (b) low to high strain rate switching. Red vertical lines are markings of the strain rate switching times. Error bars are fitting errors.}
\label{SIF3}
\end{center}
\end{figure}

Here, we have plotted the non-exponentiality fitting parameter $\beta$, used in the Kohlrausch-Williams-Watts (KWW) stretched exponential function [$r(t') = r(0)exp (-(t/\tau_{seg})^\beta)$] fitting to the measured anisotropy decay curves during our strain rate switching experiments. $\beta$ has limits $0< \beta <1$ where the smaller $\beta$ indicates a broad distribution of segmental correlation times. Before each deformation experiment, we have a wide distribution of correlation times ($\beta \sim 0.4$) to start with (Fig.~\ref{SIF3}). Then under deformation $\beta$ increases as yielding occurs and reaches a steady state depending on the applied strain rate. A higher applied strain rate results in a higher steady $\beta$ value, meaning a narrower distribution of correlation times.

\clearpage
\section*{Comparison of the anisotropy decay curves recorded immediately after switching for both types of strain rate switching experiments}
\begin{figure}[!h]
\begin{center}
\includegraphics[width=1.0\textwidth]{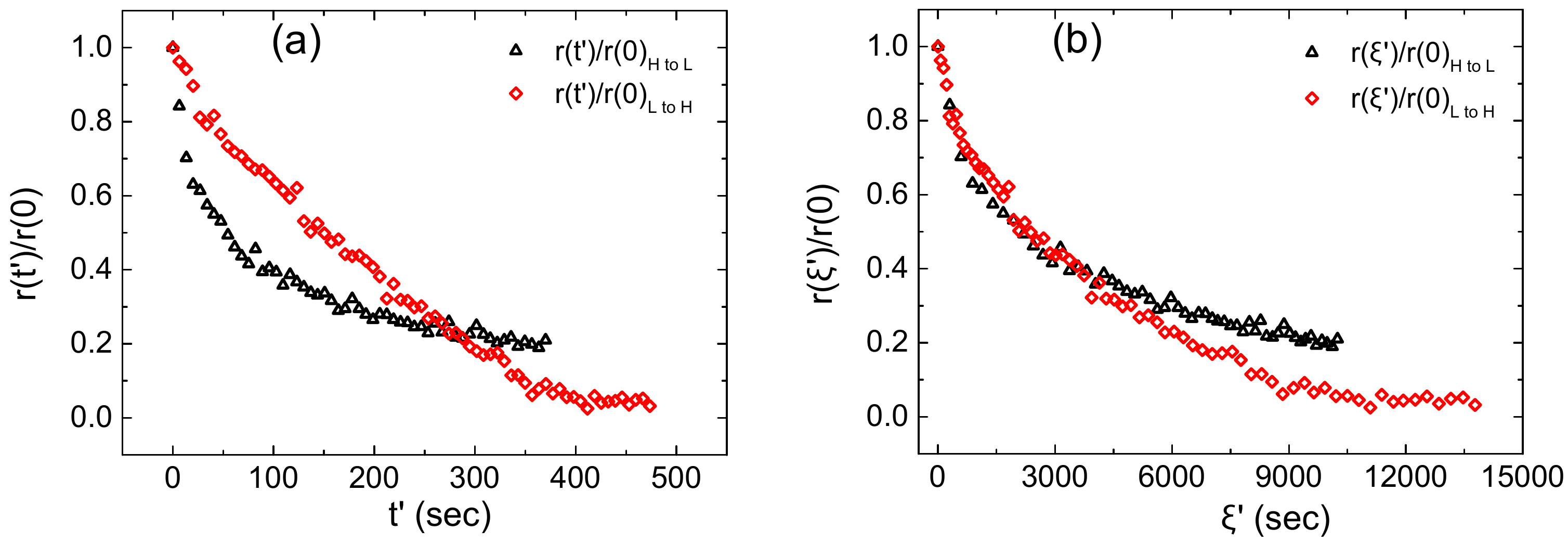}\\
\caption{(a) Anisotropy decay curves immediately after switching for both high to low and low to high strain rate switching (indicated by `H to L' and `L to H' respectively) are plotted in real time. (b) The same decay curves are re-plotted in terms of material time.}
\label{SIF4}
\end{center}
\end{figure}

Here, we have compared the anisotropy decay curves recorded immediately after switching for both types of strain rate switching. This is the most severe test of the material time assumption. First, we have plotted the two decay curves together in real time (Fig.~\ref{SIF4}a). Then we re-plotted the same anisotropy decay curves in terms of the material time (Fig.~\ref{SIF4}b). In both types of strain rate switching, we can observe a shape change due to the nonlinear relationship between experiment time ($t'$) and material time ($\xi'$). The very different anisotropy decay curves show a good overlap with each other in material time until $\xi' \approx 5600 s$ (where $r(t)/r(0) \approx 0.3$). These observations highlight the extent to which the material time concept can describe our data.

\clearpage
\section*{Characteristic time scale of variation of $\sigma$, $\tau_{seg}$, and $S_0$ after high to low switching in the Chen-Schweizer model}
\begin{figure}
\begin{center}
\includegraphics[width=0.7\textwidth]{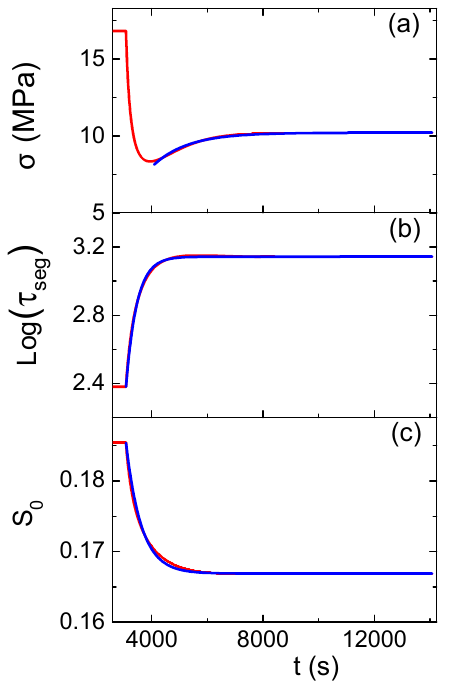}\\
\caption{We have fitted the Chen-Schweizer model output data after 3075 s (main text Fig. 4 (a-c)) with the functional form $Y = Y_s + (Y_i - Y_s)exp(-(t-t_0)/\tau_Y)$, where $t_0$ is the start of the fitting interval. (a) For $\sigma$, we have extracted the time scale to reach the steady-state from the minima $\tau_{\sigma} = 1233~s$. In (b) and (c) we have fitted the steady-state to steady-state transitions of Log($\tau_{seg}$), and $S_0$ to extract time scales of variations, $\tau_{\tau-seg} = 387~s$, and $\tau_{S} = 551~s$ respectively. Blue curves are the fittings.}
\label{SIF5}
\end{center}
\end{figure}

Here we have compared the characteristic time scales of variation of $\sigma$, $\tau_{seg}$, and $S_0$ especially after high to low strain rate switching. To extract the steady-state to steady-state transition time scales of $\tau_{seg}$ and $S_0$ in case of high to low strain rate switching, we have considered the Chen-Schweizer model output data after 3075 s in main text Fig. 4 (b-c). The data is fitted with the functional form $Y = Y_s + (Y_i - Y_s)exp(-(t-t_0)/\tau_Y)$, where $t_0~s$ is the starting time of the fitting. As $\sigma$ does not show a monotonic transition (main text Fig. 4 (a)), we have used the same functional form to extract the time to reach the steady state after the minima i.e. used $ t_0 = 4000~s$. We can see that the time scales associated with $\tau_{seg}$ and $S_0$ i.e. $\tau_{\tau-seg}$ and $\tau_{S}$ are similar, confirming that the variations of $\tau_{seg}$ and $S_0$ are coupled (Fig.~\ref{SIF5}). Further, we can observe that for Chen-Schweizer model the minima of $\sigma$ is close to where both $\tau_{seg}$ and $S_0$ become nearly steady and $\sigma$ starts to increase after that. With the experimental data in the main text Fig 1(a),(d), we have similar observations for the $\tau_{seg}$ and $\sigma$. These observations are consistent with that there is a lag between the structural variation and the change in the stress.

\clearpage
\section*{Origin of the undershoot and the overshoot in the stress after the strain rate switching in the Chen-Schweizer model}
\begin{figure}[!h]
	\begin{center}
		\includegraphics[width=1.0\textwidth]{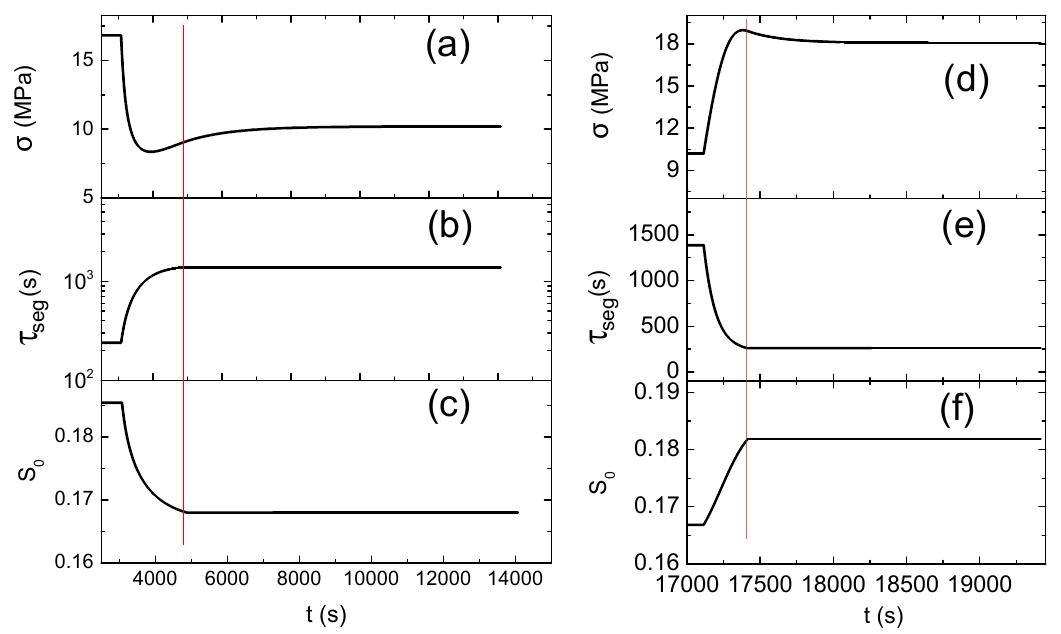}\\
		\caption{Chen-Schweizer model predictions for the variation of $\sigma$ even after $\tau_{seg}$ and $S_0$ become constant after the strain rate switching (panels a-c for high to low strain rate switching, and panels d-f for low to high strain rate switching). We have kept $\tau_{seg}$ and $S_0$ fixed after the time where $\tau_{seg}$ becomes nearly constant as indicated by the red vertical lines, and allow the $\sigma$ to vary.}
		\label{SIF6}
	\end{center}
\end{figure}

To investigate the origin of the undershoot and the overshoot in the stress after strain rate switching in the Chen-Schweizer model, we have kept the $\tau_{seg}$ and $S_0$ fixed after the $\tau_{seg}$ reaches close to the steady value, and allow $\sigma$ to vary according to the equation (1). According to the equation (1), variation of $\sigma$ is mainly controlled by the second term i.e. $-\frac{\sigma}{\tau_{seg}}$ as $\sigma << \sigma_c$. Thus, we observe the undershoot or the overshoot in $\sigma$ as the $\sigma$ requires more time to be steady even after the $\tau_{seg}$ becomes nearly constant (Fig.~\ref{SIF6}).

\end{document}